\documentclass[aps,prl,showpacs,floatfix]{revtex4}
\usepackage{graphics}
\usepackage{epsfig}
\usepackage{subfigure}
\usepackage{amsmath,amsfonts,amssymb,graphicx}

\begin{document}
\title{Supervised secure entanglement sharing for faithful quantum teleportation via tripartite W states}
\author{Yue Li}\email{liyue@smail.hust.edu.cn}
\affiliation{School of Computer Science and Technology and\\
Department of Electronics and Information Engineering, Huazhong
University of Science and Technology, Wuhan 430074, China}
\author{Yu Liu}\email{liuyu@hust.edu.cn}
\affiliation{
Department of Electronics and Information Engineering and \\
Institute of National Defense Science and Technology, Huazhong
University of Science and Technology, Wuhan 430074, China}
\date{November 18, 2007}
\begin{abstract}
We present a supervised secure entanglement sharing protocol via
tripartite W states for faithful quantum teleportation. By
guaranteeing a secure entanglement distribution in the charge of a
third believed supervisor, quantum information of an unknown state
of a 2-level particle can be faithfully teleported from the sender
to the remote receiver via the Bell states distilled from the
tripartite W states. We emphasize that reliable teleportation after
our protocol between two communication parties depends on the
agreement of the supervisor to cooperate via taking the W states as
both the quantum channel and eavesdropping detector. The security
against typical individual eavesdropping attacks is proved and its
experimental feasibility is briefly illustrated.
\end{abstract}
 \maketitle Quantum teleportation
(QT)\cite{QT} turns out to be incredibly successful in a wide
variety of quantum information science, it achieves a different way
for quantum state transmission, where arbitrary unknown quantum
state collapses on the sender side and reborns on the receiver side
via one way local operation and classical communication(LOCC) and
previously shared pairs of maximal entanglement states.

To experimentally implement the QT, however, practical obstacles
will be encountered. One is the \emph{decoherence} and the
\emph{absorption} happen in the imperfect quantum
channel\cite{decoherence}, in which particles from entanglement
pairs may get lost or entangled with the environment thus the
entanglement decreases exponentially with the extension of the
distribution distance. To address this, Zukowski \textit{et al.} put
forward the \emph{entanglement swapping} to shorten the direct
entanglement distribution\cite{Zuko}, Bennett \textit{et al.}
introduced the \emph{entanglement distillation} to refine the
quality of entanglement effected by the decoherence after
transmission\cite{decoherence}. Based on these two contributions,
\emph{quantum repeater}(QR), which was proposed by Briegel
\textit{et al.}, was constructed for long distance entanglement
distribution in noisy and lossy quantum channel\cite{Briegel}, the
invention of QR makes long distance entanglement distribution
possible and because the maximal entanglement pairs can be distilled
from the imperfect entanglement ones, the assumption that
entanglement distributions happen in a noise free environment
becomes reasonable for QT-based quantum communication protocol
design.

Another considerable obstacle is the security --- \emph{the
potential eavesdropping}. Recently, various quantum communication
protocols have been proposed based on the QT, most of which,
however, are constructed under the popular hypothesis that the well
known ``third trustable party"(TTP) distributes the particles to the
communication attendants without any interruption from the
eavesdropper who may want to stole the sender's quantum information
by recovering the qubits at her own hand, therefore, the protocol
may not work well or even fail when the quantum channels are
unauthenticated.

To be more general, the QT requires a secure enough quantum channel
--- for the direct quantum transmission to achieve the previous
sharing of the entanglements --- yet available channels are
typically with evil quantum scientist who owns unlimited computing
power. Through such considerations, we propose a supervised secure
entanglement sharing protocol, the ``Wuhan'' protocol, for QT, in
which a third supervisor Charlie is included to take in charge of
the entanglement distribution and eavesdropping detection with
classical communication and sequence of tripartite W
states\cite{Dur}. Under the help of this protocol, quantum
information of an unknown state of a 2-level particle can be
faithfully teleported.

\begin{figure}[htbp]
\begin{center}
\begin{tabular}{c}
\includegraphics[height=5cm]{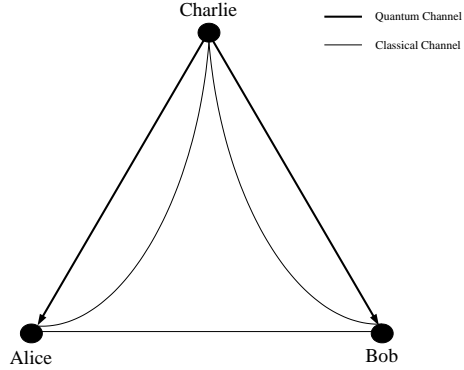}
\end{tabular}
\end{center}
\caption[example]
{ \label{fig:protocol} The scenario of the ``Wuhan" protocol. }
\end{figure}

The ``Wuhan'' protocol (see Fig.~\ref{fig:protocol}) enables
faithful QT between Alice and Bob only when the TTP Charlie agrees
to provide them the needed Bell states. The interesting scenario
perfectly suits the bridge building rule in the Wuhan city. If a new
Yangtse River bridge was going to be built between Wuchang(WC) and
Hanyang(HY), two of the three member towns of the Wuhan city, should
be approved by the governing town Hankou(HK) after which HK sends
hardhats to WC and HY using HK-WC and HK-HY bridges respectively.
Back from the tale, in this protocol we ask a believed supervisor
Charlie to help the two parties Alice and Bob for distilling the
``hardhat'' --- the Bell state which is nondeterministically
``hidden'' in the following tripartite W state
\begin{equation}
\begin{split}
|W\rangle&=\frac{1}{\sqrt{3}}(|100\rangle+|010\rangle+|001\rangle)_{abc}\\
&=\sqrt{\frac{2}{3}}[\frac{1}{\sqrt{2}}(|10\rangle+|01\rangle)_{ab}]|0\rangle_{c}+\frac{1}{\sqrt{3}}|00\rangle_{ab}|1\rangle_{c}\\
&=\frac{1}{\sqrt{3}}(|++\rangle-|--\rangle)_{ab}|0\rangle_{c}+\frac{1}{2\sqrt{3}}(|++\rangle+|+-\rangle+|-+\rangle+|--\rangle)_{ab}|1\rangle_{c},
\end{split}
\end{equation}
where $|+\rangle=\frac{1}{\sqrt{2}}(|0\rangle+|1\rangle)$ and
$|-\rangle=\frac{1}{\sqrt{2}}(|0\rangle-|1\rangle)$. Note that when
measuring either of the three qubits, say particle \textit{c}, Bell
state
\begin{equation}
|\psi^{+}\rangle=\frac{1}{\sqrt{2}}(|01\rangle+|10\rangle),
\end{equation}
made up of particles $a$ and $b$ can be obtained with a probability
$P(|\psi^{+}\rangle)= \frac{2}{3}$, then the two communication
parties cooperate with Charlie to execute two fundamental
progresses, the \emph{entanglement distribution}(ED) and
\emph{teleportation confirmation}(TC), in sequence. The ED helps to
perform secure distribution of entanglement states to Alice and Bob;
the TC, if Charlie permits, distills the ``hidden'' entanglement
states for Alice and Bob by Charlie's local measurements. Here we
give an explicit algorithm for ``Wuhan'' protocol, after which its
security is discussed by examing several typical attacks.

(ED. 0) Protocol is initialized, $t=0$. Charlie announces the switch
to \emph{transmission mode} and prepares three-qubit W states in EQ.
1, forming the sequence  $w^{N}=(w_{1},w_{2},...,w_{i},...,w_{n})$,
where $w_{i}=(a_{i},b_{i},c_{i})$ are the three qubits of the $i$th
W state. Here we assume the quantum state that Alice wants to
teleport is
\begin{equation}
|\psi\rangle_{m}=a|0\rangle+b|1\rangle,
\end{equation}
where $|a|^{2}+|b|^{2}=1$.

(ED. 1) $t=t+1$. Charlie reserves qubit $c_{t}$ from $w_{t}$ as
\emph{home qubit}, sending the other two particles $a_{t}$ and
$b_{t}$ to Alice and Bob respectively as the \emph{travel qubits}.
Repeat step (ED.1) until $t$ reaches $n+1$. As result, Charlie
obtains a \emph{home qubit} sequence
$c^{N}=(c_{1},c_{2},...,c_{i},...,c_{n})$, and the two communicators
are expected to obtain \emph{travel sequences}
$a^{N}=(a_{1},a_{2},...,a_{i},...,a_{n})$ and
$b^{N}=(b_{1},b_{2},...,b_{i},...,b_{n})$ respectively.

(ED. 2) Charlie announces the switch to \emph{detecting mode},
initializing $m=1$. Starting from his particle $c_{m}$ in $c^{N}$,
Charlie performs a nondeterministic single qubit measurement using
basis $B_{z}=\{(|0\rangle,|1\rangle)\}$, that is, the particle
$c_{m}$ owns a probability $d$ to be measured and the whole W state
collapses(also, $1-d$ to be passed by without any operation).
$m=m+1$, repeat (ED. 2) until $m=n+1$. Charlie takes down the
locations of the particles measured, forming the location
information sequence $k^{V}=(k_{1}, k_{2},..., k_{i},..., k_{v})$
and related result sequence $Rc^{V}=(Rc_{1}, Rc_{2},..., Rc_{i},...,
Rc_{v})$ where $k_{i}$ is the location of the $i$th measured
particle in $c^{N}$ satisfying $max(k^{V}) \le n$, and $Rc_{i}$ is
the measurement result for the \emph{home qubit} $c_{k_{i}}$. Also,
Charlie selects $v$ local measurement basises, each of which is in
the $B_{x}$ with probability $p$  and $B_{x}$ with $1-p$, then he
integrates the $k^{V}$ and the basises into the sequence
$kb^{V}=((k_{1},b_{1}), (k_{2},b_{2}),..., (k_{i},b_{i}),...,
(k_{v},b_{v}))$ where $k_{i}\in k^{V}$ and $b_{i} \in
\{B_{x},B_{z}\}$, he then sends the $kb^{V}$ to the classical
channel, keeping the sequence $Rc^{V}$ himself.

(ED.~3) Following the $kb^{V}$, Alice and Bob both perform local
measurements on their own \emph{travel qubit} sequences using the
combined basis in $kb^{V}$  , giving two measurement result
sequences $Ra^{V}=(Ra_{1}, Ra_{2},..., Ra_{i},..., Ra_{v})$ for
Alice and $Rb^{V}=(Rb_{1}, Rb_{2},..., Rb_{i},..., Rb_{v})$ for Bob.
Unlike Charlie's keeping on $Rc^{V}$ in Step (ED. 2), they send the
two result sequences to the public classical channel.

(ED.~4) On receiving the two result sequences, Charlie performs the
following checking algorithm for all elements of the two sequences:

\begin{table}[h]
\begin{tabular}{|p{17cm}|}
\hline\textsf{(Alg.~0) If $b_{i}=B_{z}$, let $\mathbb{C}[0]$ and
$\mathbb{C}[1]$ be the two checking flags for $Ra^{V}$ and $Rb^{V}$
respectively;}\\

\textsf{(Alg.~1) $\mathbb{C}[0]=0$ if $Ra_{i} \oplus Rb_{i}=0$ for
all $Rc_{i}$ s which are $|0\rangle$ in $Rc^{V}$, else
$\mathbb{C}[0]=1$;}\\

\textsf{(Alg.~2) $\mathbb{C}[1]=0$ if $Ra_{i} \oplus Rb_{i}=1$ for
all $Rc_{i}$ s which are $|1\rangle$ in $Rc^{V}_{d}$, else
$\mathbb{C}[1]=1$;}\\

\textsf{(Alg.~3) $(\mathbb{C}[0]\&\&\mathbb{C}[1])=0$: Eve is
detected, Charlie announces the abortion of this  distribution and
goes to step (ED.~0); $(\mathbb{C}[0]\&\&\mathbb{C}[1])=1$: proceeds
(TC.~0);}\\

\textsf{(Alg.~4) Else, let $\mathbb{C}[2]$ be the checking flag;}\\

\textsf{(Alg.~5) $\mathbb{C}[2]=0$ if $Ra_{i} \neq Rb_{i}$ for all
$Rc_{i}$ s which are $|0\rangle$ in $Rc^{V}$, else
$\mathbb{C}[2]=1$;}\\

\textsf{(Alg.~6) $\mathbb{C}[2]=0$: Eve is detected, Charlie announces the
abortion of this  distribution and goes to step (ED. 0);
$\mathbb{C}[2]=1$: proceeds (TC. 0).}\\
\hline
\end{tabular}
\end{table}

(TC.~0) Charlie ticks out all his measured \emph{home qubits},
leaving the new \emph{home qubit} sequence $c'^{N-V}=(c'_{1},
c'_{2},...,c'_{i},...,c'_{n-v})$ which are particles not selected in
the \emph{detecting mode}, if Charlie agrees with Alice and Bob's QT
requirement, he performs local measurements on all the \emph{home
qubits} in $c'^{N-V}$, obtaining result sequence
$Rc'^{N-V}=(Rc'_{1}, Rc'_{2},..., Rc'_{i},..., Rc'_{n-v})$. Note
that after his measurements, the joint state of particles $m$, $a$
and $b$ may become one of the following states
\begin{align}
\begin{split}
|\phi_{0}\rangle &=|\psi\rangle_{m}\otimes \ _{c}\langle0|W\rangle_{abc} \\
               &=(a|0\rangle+b|1\rangle)_{m}\otimes|00\rangle_{ab} \\
\end{split}
\\
\begin{split}
|\phi_{1}\rangle &=|\psi\rangle_{m}\otimes \ _{c}\langle1|W\rangle_{abc}\\
                       &=(a|0\rangle+b|1\rangle)_{m}\otimes\frac{1}{\sqrt{2}}(|10\rangle+|01\rangle)_{ab},\\
\end{split}
\end{align}
Charlie runs a selection to produce another location information
sequence $k'^{S}=(k'_{1}, k'_{2},..., k'_{i},..., k'_{s})$
satisfying $max(k'^{S}) \le (n-v)$ and $Rc'_{k'_{i}}$ in $Rc'^{N-V}$
is $|0\rangle$, then he publishes the sequence $k'^{V}$.

(TC.~1) According to $k'^{V}$, Alice and Bob locate and extract
their own particles from  $a^{N}$ and $b^{N}$, taking particles
$a_{k'_{i}},b_{k'_{i}}$ as a whole, maximal entanglement pairs in
Bell state $|\phi^{+}\rangle$ are successfully distilled and the
required previous sharing is achieved, therefore further faithful QT
between Alice and Bob can be pushed forward.

\begin{figure}[htbp]
\begin{center}
\begin{tabular}{c}
\includegraphics[height=5cm]{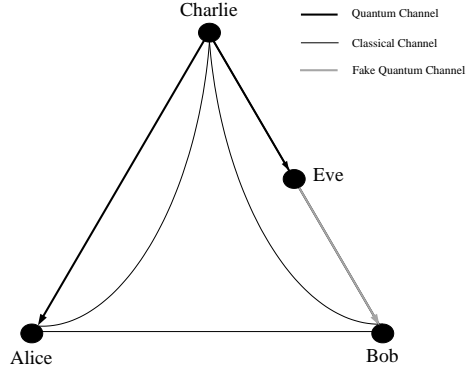}
\end{tabular}
\end{center}
\caption[example]
{ \label{fig:IRA} Eavesdropping by the intercept-resend attack. }
\end{figure}

We now consider the security of the protocol and first take account
of the intercept attack. Suppose the direction of the final QT are
from Alice to Bob, it is sure that if Eve proposes only naive
interceptions without appending fake particles, she may soon be
detected when the three parties do the particle number comparison,
therefore one considerable approach for Eve is the
\emph{intercept-resend attack} (IRA, see Fig.~\ref{fig:IRA}) which
may include two different strategies, the
\emph{intercept-measure-resend attack} (IMRA) and the
\emph{intercept-store-resend attack} (ISRA), here we show the
robustness of our protocol in case of these two IRA.

In the IMRA, Eve intercepts the particle which is going to the Bob
side from Charlie and performs immediate single qubit measurement,
then she appends fake particle which is identical with the
measurement result to Bob which establishes fake quantum channel
between Eve and Bob. To be more specific, because the initial state
is $|W\rangle$, the evaesdropping may turns out to be in state
$|0\rangle$ with probability $P(0)=\frac{2}{3}$, she then sends a
fake particle in state $|0\rangle$ to Bob, thus joint state for
Alice, Bob and Charlie's system becomes
\begin{equation}
\begin{split}
|w_{1}\rangle=&\frac{1}{\sqrt{2}}(|10\rangle+|01\rangle)_{ac}\otimes|0\rangle_{b}\\
            =&\frac{1}{\sqrt{2}}(|100\rangle+|001\rangle)_{abc};\\
\end{split}
\end{equation}
or she may obtain state $|1\rangle$ with probability
$P(|1\rangle)=\frac{1}{3}$, sending another particle in state $|1\rangle$ to
Bob, leaving the collapsed fake state
\begin{equation}
\begin{split}
|w_{2}\rangle=&|00\rangle_{ac}\otimes|1\rangle_{b}\\
=&|010\rangle_{abc}.\\
\end{split}
\end{equation}
In either case, we conclude that Eve can be detected with
probability during the \emph{detecting mode} of this protocol,
however, we are going to ignore the calculations for these
probabilities since the principals of the QT do not allow the
collapsed but unentangled states  $|w_{1}\rangle$ and
$|w_{2}\rangle$ being the appropriate sources between Eve and Alice
to recover Alice's quantum states, in other words, the IMRA can not
bring Eve any quantum information.

Coming to the other IRA strategy, the ISRA, where Eve protects the
intercepted particle by just storing without measurement to wait for
a later impersonating teleportation. In order to hide her existence,
Eve then sends a fake qubit from herself to Bob, different from the
IMRA, Eve does not know the exact state of the particle, thus she
can only send a particle in state
$|\phi\rangle_{b(e)}=(x|0\rangle+y|1\rangle)_{b(e)}$ in which $x$,
$y$ are random coefficients satisfying $x^{2}+y^{2}=1$, then the
joint state of the quad-party system turns to
\begin{equation}
\begin{split}
|\Upsilon\rangle = &|\phi\rangle_{b}\otimes|W\rangle_{aec}\\
=&(x|0\rangle+y|1\rangle)_{b}\otimes\frac{1}{\sqrt{3}}(|100\rangle+|010\rangle+|001\rangle)_{aec}\\
=&\frac{1}{\sqrt{3}}(x|1000\rangle+y|0101\rangle)_{aecb}\\
+&\frac{1}{\sqrt{3}}(x|0100\rangle+y|1001\rangle)_{aecb}\\
+&\frac{1}{\sqrt{3}}(x|0010\rangle+y|0011\rangle)_{aecb}\\
\end{split}
\end{equation}
For convenience, we ignore the intercepted particle $e$, and write
the Eq. (8) into
\begin{equation}
\begin{split}
|\Upsilon'\rangle = &\frac{1}{\sqrt{3}}(x|100\rangle+y|010\rangle)_{abc}\\
+&\frac{1}{\sqrt{3}}(x|000\rangle+y|110\rangle)_{abc}\\
+&\frac{1}{\sqrt{3}}(x|001\rangle+y|011\rangle)_{abc},\\
\end{split}
\end{equation}
according to the checking algorithm in (ED. 4):

\textbf{Case~1.} If Charlie's measurement obtains state $|1\rangle$,
combining the basis $B_{z}$ with probability $p$, Eve may be
detected due to the introduced state
$|Rej_{1}\rangle=|011\rangle_{abc}$ with probability
\begin{equation}
P(d_{1})=\frac{1}{3}\cdot p\cdot d\cdot y^{2}=\frac{pdy^{2}}{3};
\end{equation}

\textbf{Case~2.} If Charlie obtains state $|0\rangle$ combining
$B_{z}$, Eve may be detected by the state
$|Rej_{2}\rangle=x|000\rangle+y|110\rangle$ with the probability
\begin{equation}
P(d_{2})=\frac{1}{3}\cdot p\cdot d=\frac{pd}{3};
\end{equation}

\textbf{Case~3.} If Charlie combines the basis $B_{x}$, the
eavesdropping attack succeeds no matter what Charlie's measurement
results.

\begin{figure}[htbp]
\centering \subfigure[]{
\label{fig:subfig:a} 
\includegraphics[width=2.5in]{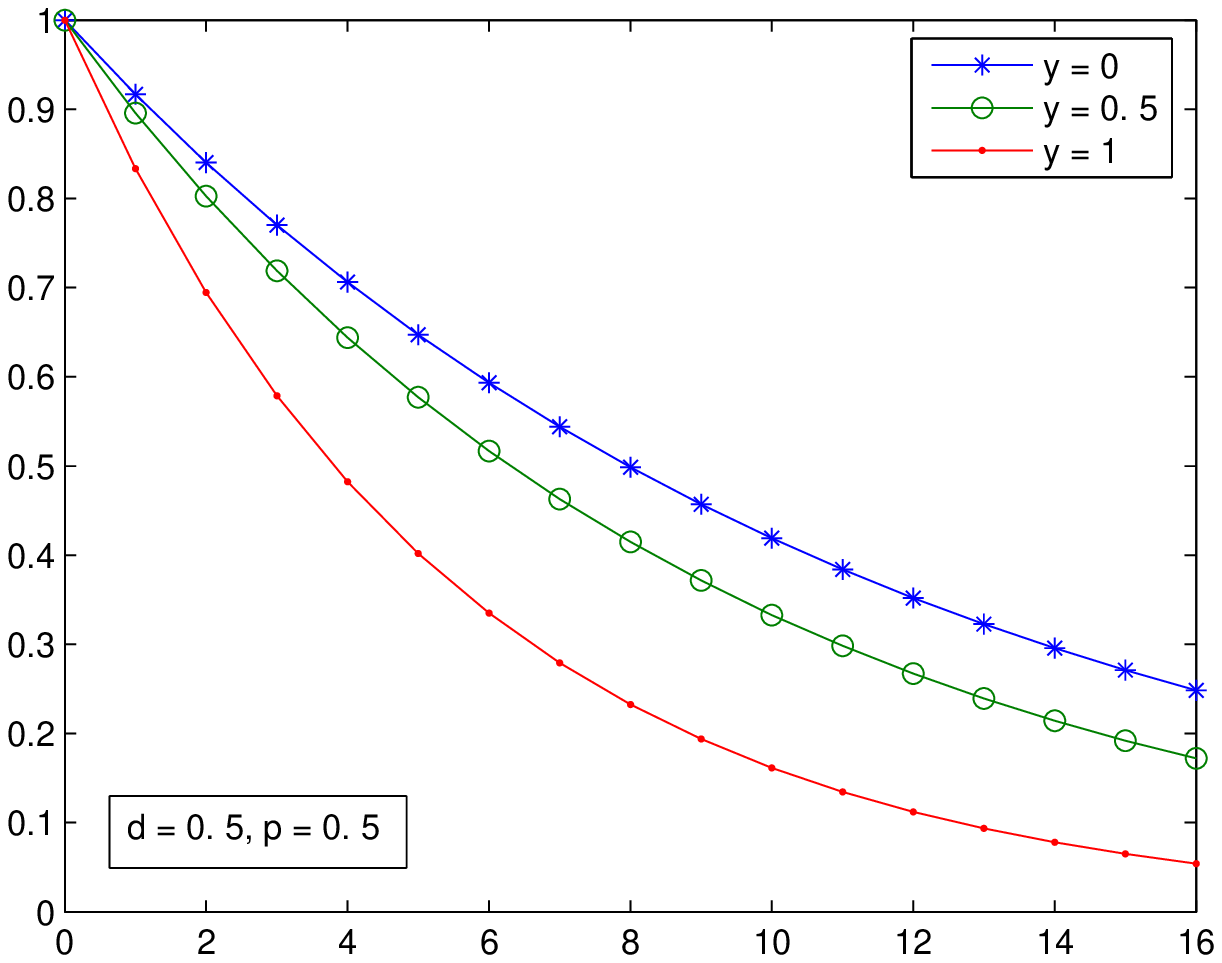}}
\hspace{0.05in} \subfigure[]{
\label{fig:subfig:b} 
\includegraphics[width=2.5in]{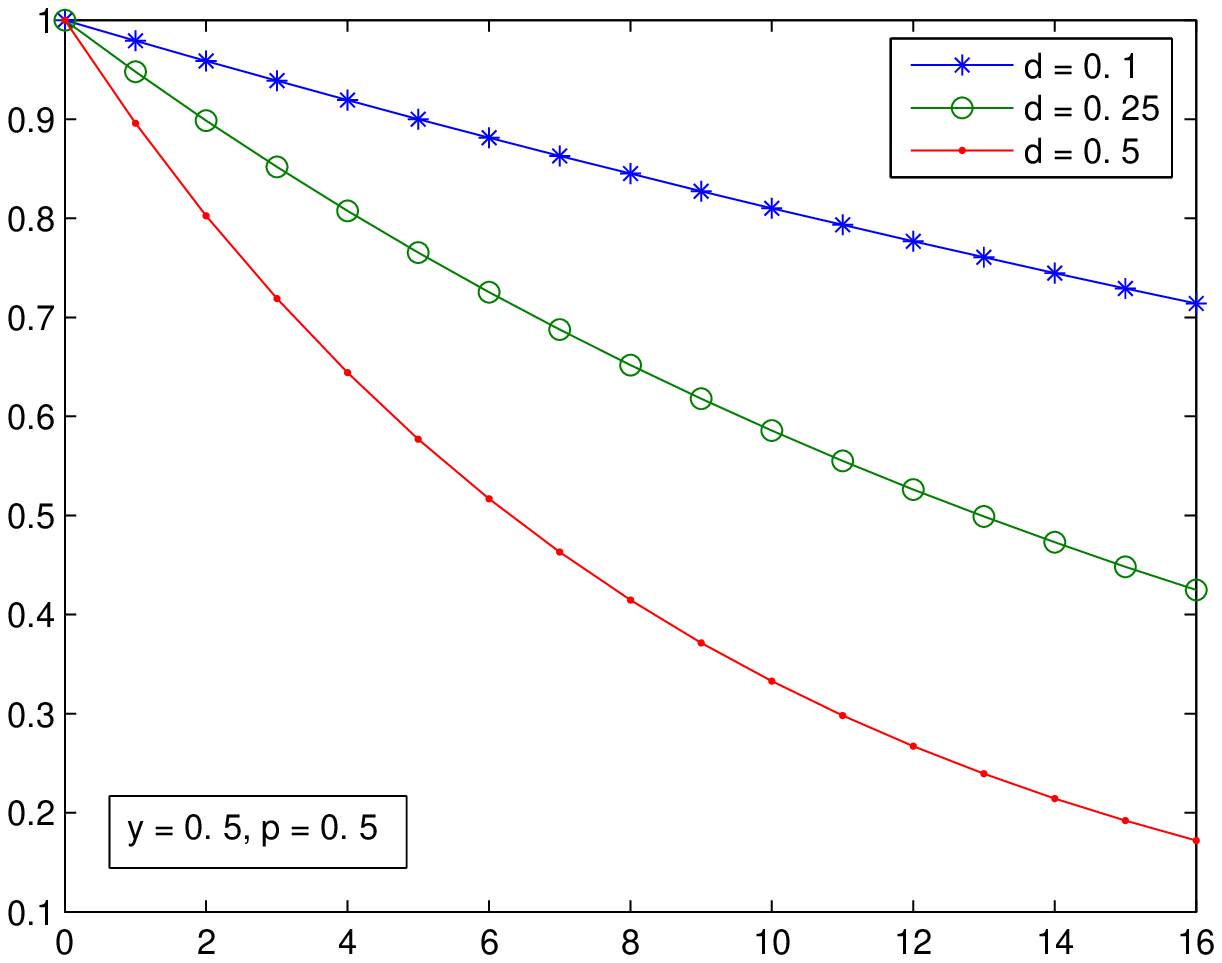}}
\hspace{0.05in} \subfigure[]{
\label{fig:subfig:c} 
\includegraphics[width=2.5in]{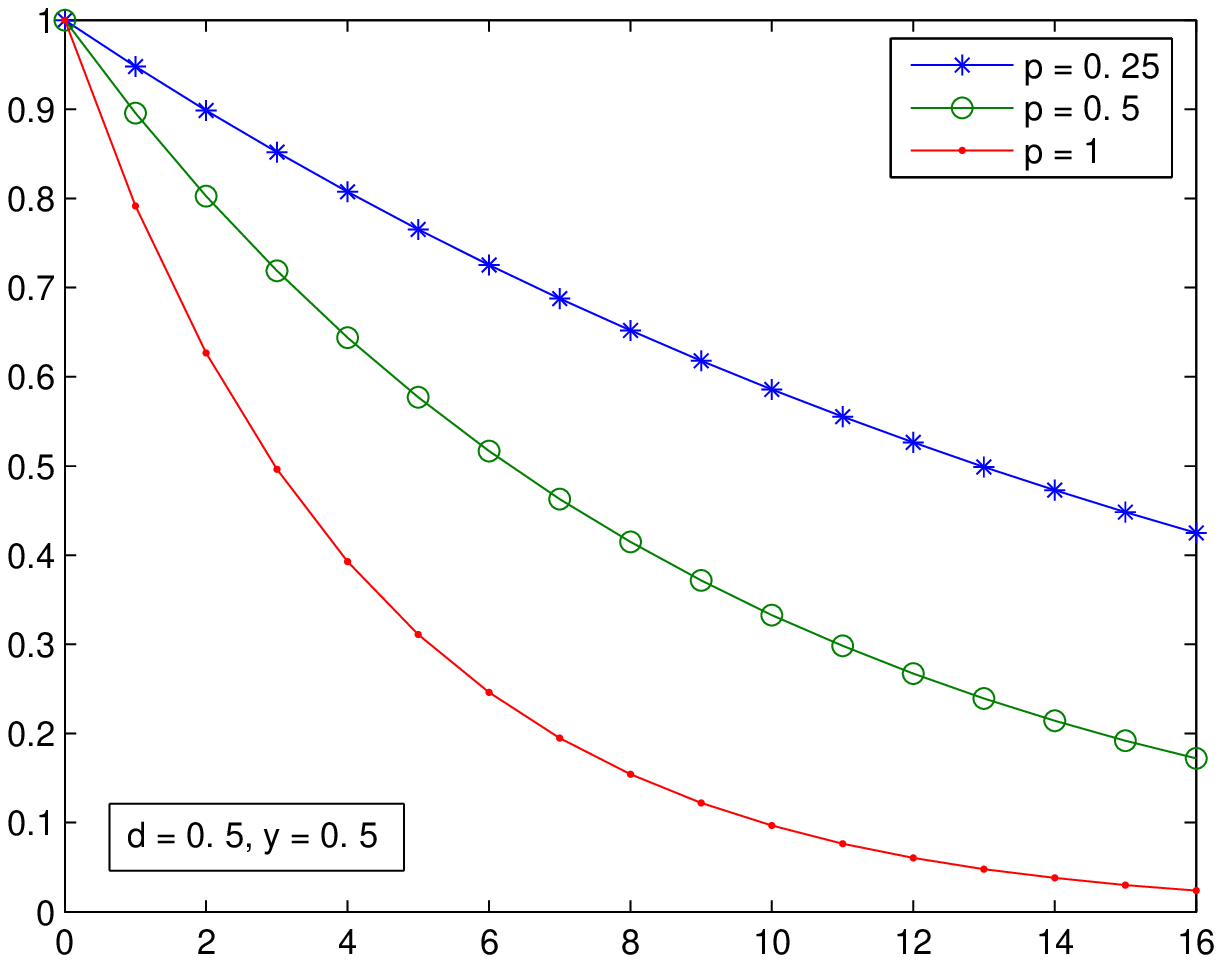}}
\caption{Successful eavesdropping rates, plotted for (a) specified
probability $d$ and $p$ with different state coefficients $y$; (b)
specified $y$ and $p$ with different $d$; (c) specified $d$ and $y$
with different $p$}
\label{fig:ProbS} 
\end{figure}

Therefore, the probability of a successful ISRA eavesdropping for
one particle is
\begin{equation}
\begin{split}
S(y,p,d,1)=&1-P(d_{1})-P(d_{2})\\
=&1-\frac{pd}{3} - \frac{pdy^{2}}{3}\\
=&1-\frac{pd(1+y^{2})}{3},
\end{split}
\end{equation}
and the rate for whole sequence $w^{N}$ is
\begin{equation}
\begin{split}
S(y,p,d,n)=&S(y,p,d,1)^{n}\\
=&[1-\frac{pd(1+y^{2})}{3}]^{n}.
\end{split}
\end{equation}
For $0 \le y,p,d \le 1$, $S(y,p,d,n)$ decreases exponentially but is
always nonzero. In the limit $n\rightarrow\infty$, we have
$S(y,p,d,n)\rightarrow0$, thus the protocol is \emph{quasi-secure}.
Here we give the curves for $S(y,p,d,n)$ in Fig.~\ref{fig:ProbS} as
examples for cases of different $y,p,d$ selections.

\begin{figure}[htbp]
\begin{center}
\begin{tabular}{c}
\includegraphics[height=5cm]{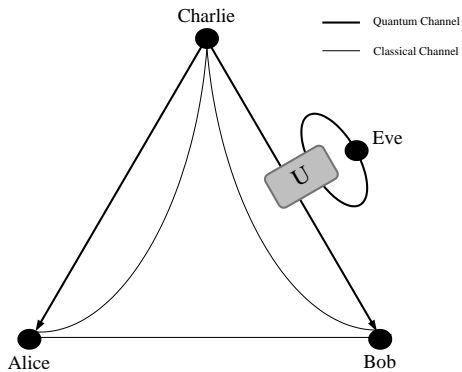}
\end{tabular}
\end{center}
\caption[example]
{ \label{fig:EMA} The entangle-measure attack. }
\end{figure}

Another typical attack different from the IRA is the
\emph{entangle-measure attack}(EMA, see Fig.~\ref{fig:EMA}), i.e.
Eve may uses her own \emph{ancillar qubit} which is in state
$|0\rangle$ to entangle with the \emph{travel qubit} $b$ being sent
to Bob by utilizing a CNOT gate $U_{be}$ (let Bob's particle be the
\emph{controller} and Eve's be the \emph{target})\cite{Nielsen},
then the joint state of the four particles becomes
\begin{equation}
U_{be}|W_{abc}0_{e}\rangle=\frac{1}{\sqrt{3}}(|1000\rangle+|0101\rangle+|0010\rangle)_{abce},
\end{equation}
since the EMA never modifies the original state of the three-qubit W
state, this kind of attack can not be sensed during the protocol,
however, Eve can still be found during the final QT, the reason is
that after Eve's CNOT operation, the extracted Bell state used for
QT actually becomes $|\phi'\rangle =
\frac{1}{\sqrt{2}}(|100\rangle+|011\rangle)_{abe}$, with the state
$|\psi\rangle_{m}$, we have the joint state
\begin{equation}
\begin{split}
|\psi\rangle_{m}|\phi'\rangle_{abe}=&(a|0\rangle+b|1\rangle)_{m}\otimes\frac{1}{\sqrt{2}}(|100\rangle+|011\rangle)_{abe}\\
=&\frac{1}{\sqrt{2}}(a|0100\rangle+a|0011\rangle+b|1100\rangle+b|1011\rangle)_{mabe}\\
=&\frac{1}{2}|\psi^{+}\rangle_{ma}|\varepsilon^{+}\rangle_{be}\\
+&\frac{1}{2}|\psi^{-}\rangle_{ma}|\varepsilon^{-}\rangle_{be}\\
+&\frac{1}{2}|\phi^{+}\rangle_{ma}|\xi^{+}\rangle_{be}\\
+&\frac{1}{2}|\phi^{-}\rangle_{ma}|\xi^{-}\rangle_{be},\\
\end{split}
\end{equation}
where
$|\psi^{\pm}\rangle=\frac{1}{\sqrt{2}}(|01\rangle\pm|10\rangle)$,$|\phi^{\pm}\rangle=\frac{1}{\sqrt{2}}(|00\rangle\pm|11\rangle)$,$|\varepsilon^{\pm}\rangle=(a|00\rangle
\pm b|11\rangle)$ and $|\xi^{\pm}\rangle=(a|11\rangle \pm
b|00\rangle)$, Eve may recover the state that Alice teleports on the
first particle of her \emph{ancillar qubit} sequence, however if Eve
continues to perform measurement for specifying the coefficients $a$
and $b$, the entanglement state made up from particles $b$ and $e$
collapses, therefore Bob's measurement on his particle thus only
gives deterministic result, which implies Eve's existence.

The ``Wuhan''  protocol is experimentally feasible. The tripartite W
states can be produced by the parametric down-conversion
method\cite{Ekert}; the storage of photons is necessary only for a
duration corresponding to the distance of Charlie's distributions
and the quantity of qubits that Alice wants to teleport and the
entanglement sharing efficiency depends on Charlie's checking
parameters $d$ and $p$; the local measurements and the QT have been
already experimentally implemented. Altogether, the realization of
the ``Wuhan'' protocol should be reachable using current quantum
information technology.

\appendix
\acknowledgments Y.~Li would like to thank Qing-yu Cai for fruitful
discussion. This research is supported by the Innovation Foundation
of Aerospace Science and Technology of the China Aerospace Science
and Technology Corporation (CASC) under Grant No.~20060110 and the
National Foundation for Undergraduate Novel Research from the
Ministry of Education of P.~R.~China.

\end{document}